\documentclass[10pt,twocolumn,letterpaper]{article}

\usepackage{cvpr}
\usepackage{times}
\usepackage{epsfig}
\usepackage{graphicx}
\usepackage{amsmath}
\usepackage{amssymb}
\usepackage{url}
\usepackage[shortlabels]{enumitem}

\usepackage[breaklinks=true,bookmarks=false]{hyperref}

\cvprfinalcopy 

\def\cvprPaperID{1345} 

\begin{document}

\title{US Fatal Police Shooting Analysis and Prediction}

\author{Yuan Wang\\
University of Rochester\\
{\tt\small yuan.wang1@simon.rochester.edu}
\and
Yangxin Fan\\
University of Rochester\\
{\tt\small yfan24@ur.rochester.edu}
}

\maketitle
\begin{abstract}
We believe that “all men are created equal” \cite{1776Thomas}. With the rise of the police shootings reported by media, more people in the U.S. think that police use excessive force during law enforcement, especially to a specific group of people. We want to add our two cents point of view by multi-dimensional analysis to reveal more facts than the monotone mainstream media. The more facets we understand the problem, the better solution that the whole society could approach.  
\end{abstract}

\section{Introduction}
Our report has three parts: First, we analyzed and quantified fatal police shooting news reporting deviation of mainstream media. Second, we used FP-growth to mine frequent patterns, clustered hotspots of fatal police shootings, and brought multi-attributes (social economics, demographics, political tendency, education, gun ownership rate, police training hours, etc.) to reveal connections under the iceberg. Third, we built regression models based on correlation analysis for numeric variables selection to predict police shooting rates at the state level. We also built classification models based on Chi-square testing for categorical variables selection to predict the victims' race of fatal police shootings. The main datasets we choose for our analysis include: 
1. Washington Post Fatal Police Shooting Dataset (WP data) \cite{2020WashingtonPost} covers fatal police shooting from 01-01-2015 to 12-02-2020. 
2. KilledByPolice (KBP): Fatal police shooting reported in KilledByPolice website \cite{2020KBP} from 01-01-2015 to 11-04-2020. 

\section{Related work}
Several studies have been conducted based on utilizing local crime data to explain racial disparities and differences in fatal police shootings. Mentch (2020) \cite{mentch2020racial} implemented resampling procedures to take factors like local arrest demography and law enforcement density into account. He found substantially less racial disparity after accounting for local arrest demographics. On the contrary, Ross (2015) \cite{ross2015multi} built a multi-level Bayesian model to investigate the extent of racial bias in the recent shooting of civilians by police. He concluded that racial discrimination observed in police shootings is not explainable due to local-level race-specific crime rates. Noticeably, Mentch and Ross had reached contradictory conclusions. But they inspired us to use data mining and machine learning techniques to incorporate more factors rather than only crime data to understand fatal police shootings in the US better. 

\section{Methodology}
We defined \textbf{reporting deviation rate} and \textbf{total absolute reporting deviation rate} to evaluate the media's reporting bias. 

In WP dataset analysis, we used \textbf{FP-growth} and \textbf{word cloud} to reveal the frequent patterns and \textbf{DBSCAN clustering} to find fatal shooting hotspots. We also implemented \textbf{correlation analysis} to analyze correlation between multiple numeric attributes and fatal police shooting rate and tested the significance of their correlations. We used \textbf{T-test/ANOVA} to measure the significance of fatal police shooting rate by categorical attributes. 

In fatal police shooting rate prediction, we used results of correlation analysis to select numeric predictors. We constructed a series of regression models, including \textbf{Kstar}, \textbf{K-Nearest-Neighbor}, \textbf{Random Forest}, and \textbf{Linear Regression}, to predict state level's fatal police shooting rate.
We measured their performance by \textbf{ten-fold cross validation} scores. In victims’ race prediction, we used \textbf{Chi-square testing} to do \textbf{variables selection}. We built a series of classification models, including \textbf{Gradient Boosting Machine}, \textbf{Multi-class Classifier}, \textbf{Logistic Regression}, and \textbf{Naïve Bayes Classifier}, to predict the race of fatal police shooting victims. We measured their performance by \textbf{stratified five-fold cross validation} scores.  

\section{Media reporting analysis}
Since 2015, The Washington Post (WP) has created a database cataloging every fatal shooting nationwide by a police officer in the line of duty. There have been less than 1,000 people killed by police every year. The killed rate of African American people is disproportionally higher than any other race (use Black or B to distinguish with Asian or A in the following). The Figure-1 show the number of people killed by police shooting by year in national wide. Figure-2 shows the average proportion rate of each race killed by police shooting from WP’s website.

\begin{figure}[h!]
  \includegraphics[scale=0.25]{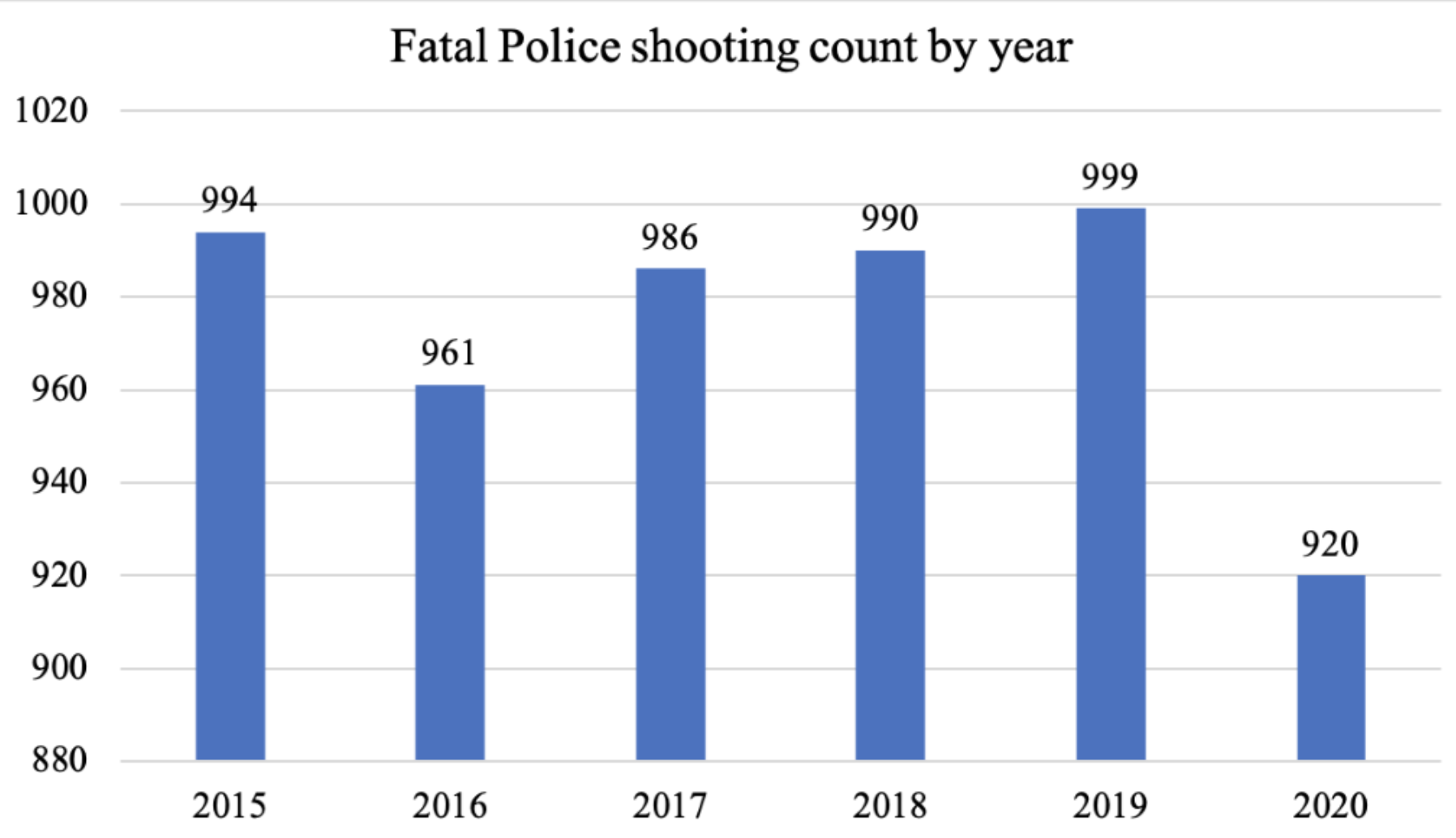}
  \caption{Number of people killed by police shooting by year till 02/12/2020}
\end{figure}

\begin{figure}[h!]
  \includegraphics[scale=0.25]{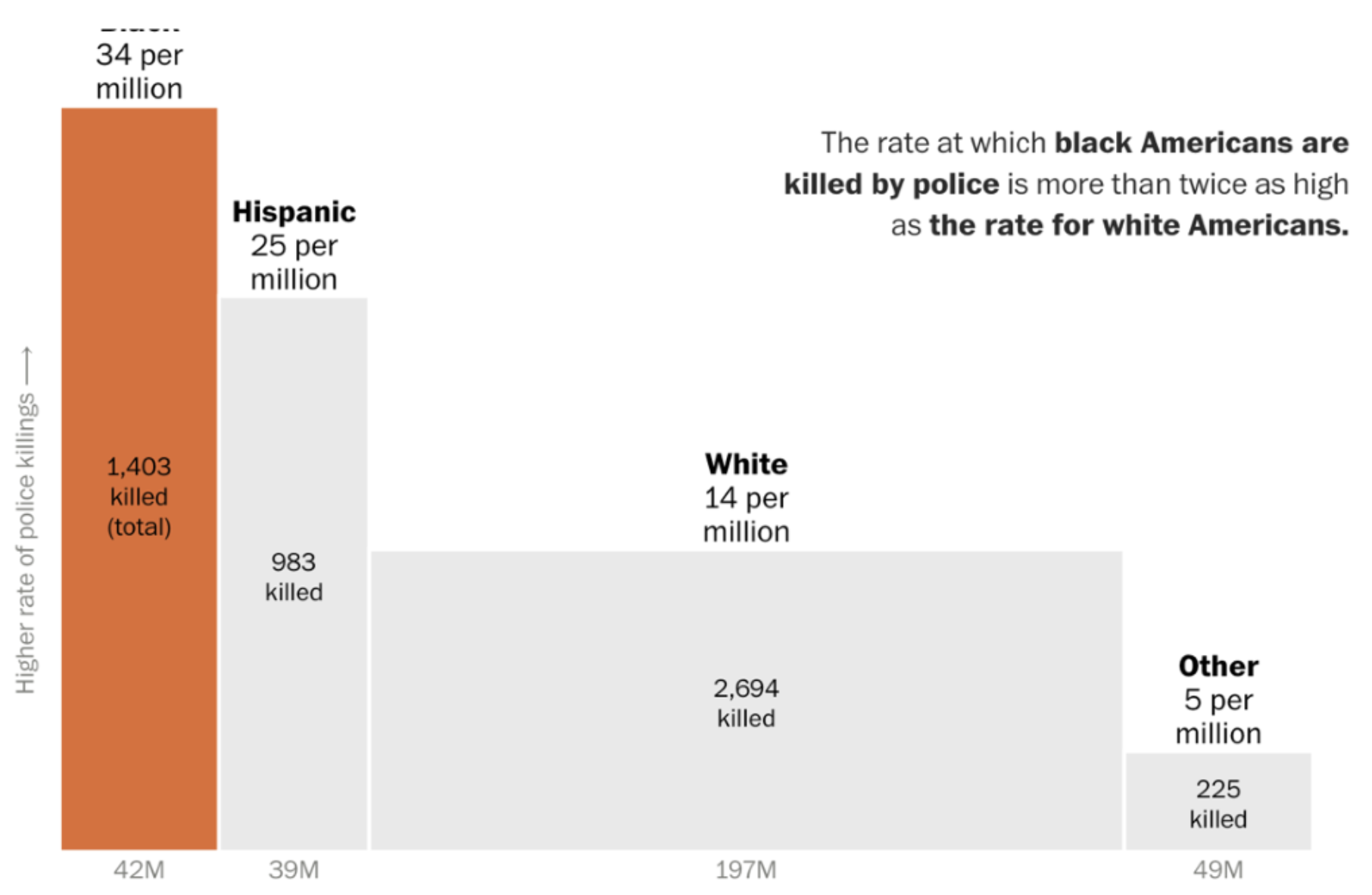}
  \caption{The average proportion rate of each race killed by police shooting, \cite{2020WP}}
\end{figure}

Admittedly, there is no doubt that Black people’s rate is high than any other group of people if we compare it with the population proportion. However, once we add the proportion of violent incidents offenders \cite{CrimeRatebyRace}to each racial group, we see the ratios have matched each other accordingly. See Figure-3 below. 

\begin{figure}[h!]
  \includegraphics[scale=0.25]{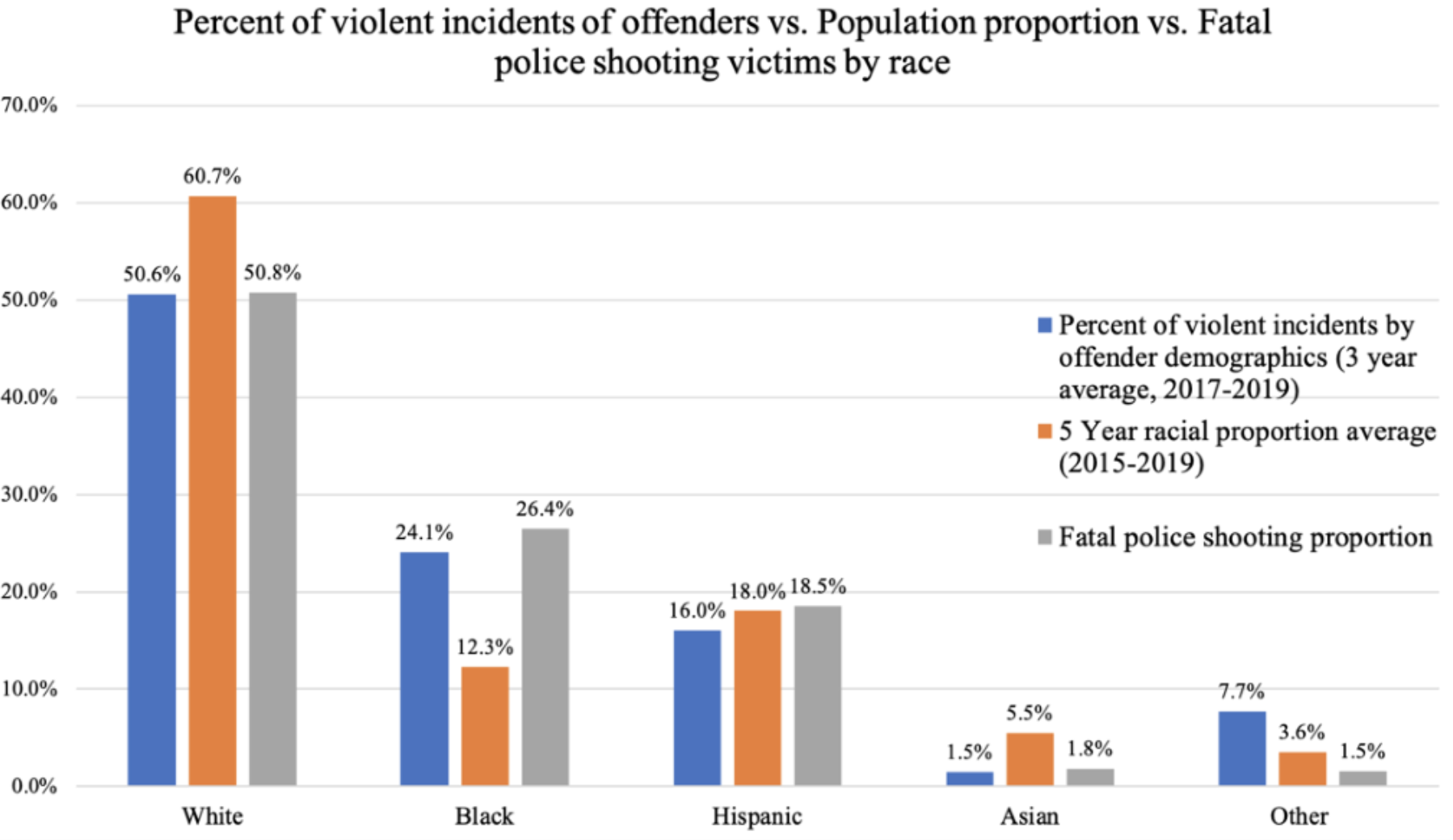}
  \caption{Percent of violent incidents of offenders (3-year average) VS. 5-year average population proportion VS. Fatal police shooting victims by race}
\end{figure}

We collected 2472 police shooting victims with known reported media and race from 2016 till now from KBP. We hold our null hypothesis that media reported news by each race should follow the real happened case distribution. We use the racial proportion of victims from WP Data as the ground truth. We selected media with over 100 fatal police shooting news reporting, which include one conservative media, FOX (318) and three liberal media: ABC (244), CBS (227), and NBC (135). The media’s political inclination is showed by Figure-4: Political bias of selected media. We excluded media with less than 100 reporting since most of them are local media whose news may be impacted by the local demographics. Figure-5 shows the comparison results: all four media have different deviations on reporting the truth. In general, Black victims were over-reported among all the media.   

\begin{figure}[h!]
  \includegraphics[scale=0.25]{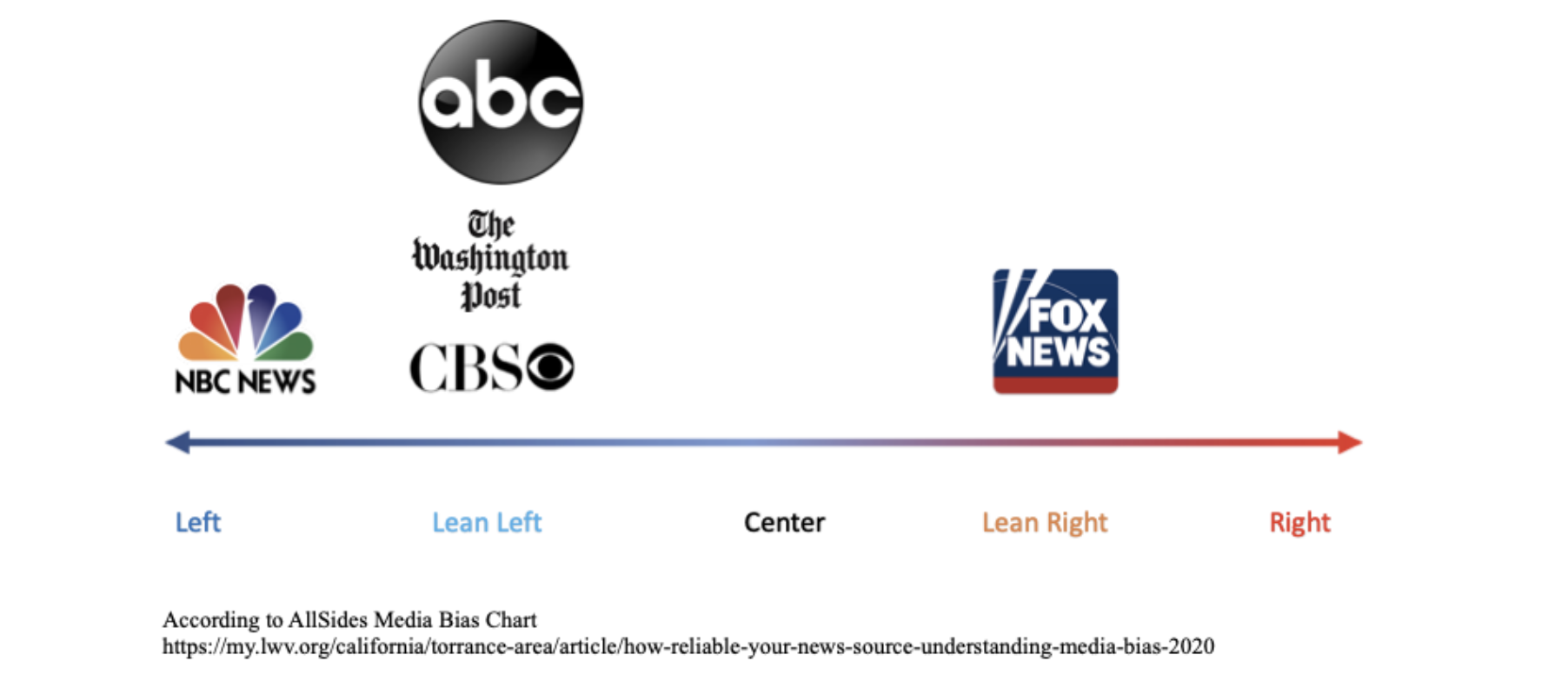}
  \caption{Political bias of selected media}
\end{figure}

\begin{figure}[h!]
  \includegraphics[scale=0.25]{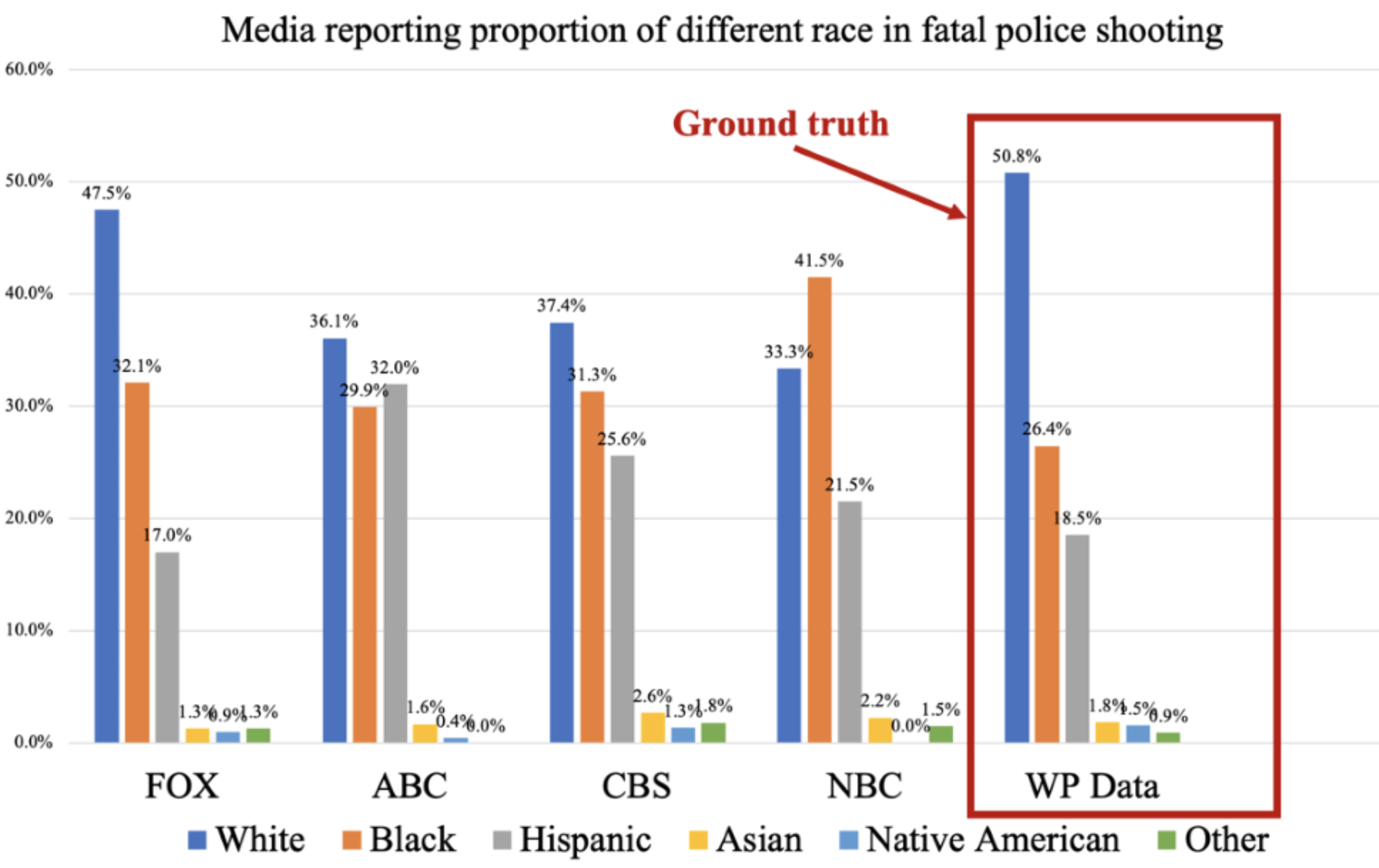}
  \caption{Media reporting proportion of police shooting by different race}
\end{figure}

To exam the difference of deviation between four media, we defined the measurements and calculation methods: 

 1. Reporting deviation rate of media B regarding race A = $R(B, A)$ = reported proportion of race A by media B – real proportion of race A in WP Data. 
 If $R(B, A) < 0$, media B underreports race A victims. 
 Else $R(B, A) > 0$, media B overreports race A victims.

 2. Total absolute reporting deviation rate of media B = $=\sum_{i=1}^{N} |R(B, A_i)|$, Ai is the i-th race and N is the number of races.

We then get Figure-6: Four media reporting deviations. FOX has the least deviation rate from the WP Data, and there are only –3.3\% deviation for White, +5.6\% for Black, -1.5\% for Hispanic, -0.6\% for Asian, and –0.6\% for Native American, and +0.4\% for Other. Nevertheless, ABC, CBS, and NBC have larger reporting deviations, which underreported 10\% White victims while overreported Hispanic and African American victims. Specifically, NBC underreported White victims' proportion by 17.4\% and overreported Black victims' proportion by 15.0\%. Furthermore, it even reported more Black victims (41.5\%) than whites (33.3\%). ABC overreported Hispanic victims' proportion by 13.4\%. It reported Hispanic victims (32.0\%) at the same level as White victims (36.1\%). The Figure-7 shows four media total absolute deviation rate. 

\begin{figure}[h!]
  \includegraphics[scale=0.25]{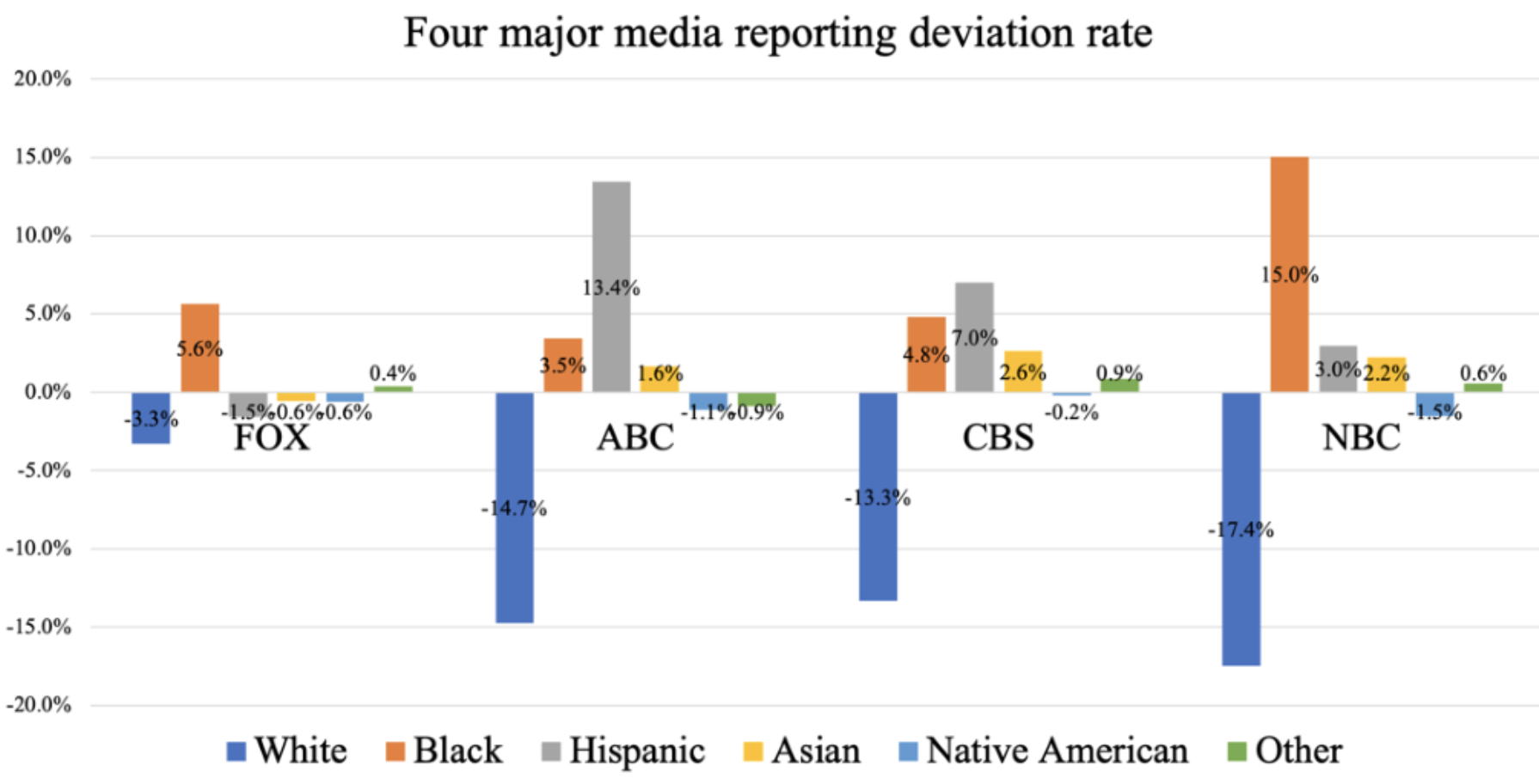}
  \caption{Four major media reporting deviation rate}
\end{figure}

\begin{figure}[h!]
  \includegraphics[scale=0.25]{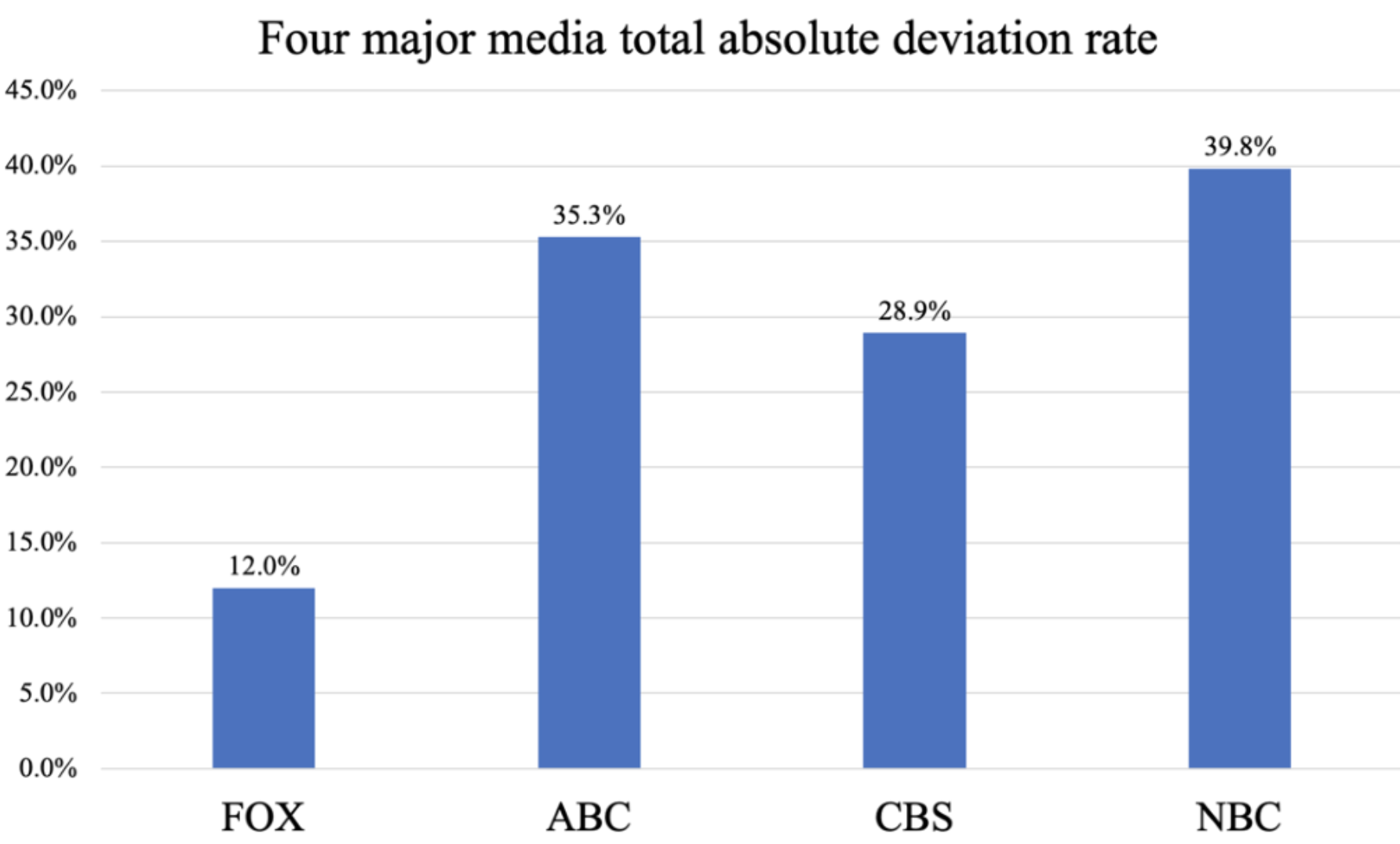}
  \caption{Four major media total absolute deviation rate }
\end{figure}

In terms of total absolute reporting proportion error, NBC has the largest reporting deviation rate (39.8\%), followed by ABC (35.3\%), and CBS (28.9\%), while FOX has the least rate (12.0\%) shown above. 
  
\section{WP fatal police shooting dataset insight}
In this part, we use FP-growth and word cloud to reveal the frequent pattern behind the WP dataset. We use location data from the WP dataset to cluster police shooting incidents and find shooting hotspots. We also tried multi-attributes such as social economics, demographics, political tendency, education, gun ownership rate, police training hours, etc., to verify the possible reason for the police shooting. 

\subsection{Frequent Pattern Mining}
From the frequent pattern mining, we can conclude a typical victim shot by police: \textbf{a “man” (96\%) “without mental illness” (77\%) uses “gun” (57\%) “attack” (65\%) police then get “shot” (95\%) by police who does not wear “body camera” (88\%)}. see below Figure-8 and Figure-9. \textbf{“California,” “Texas,” “Florida”} are the top three states were happened more frequently in total number, see Figure-10.  

Therefore, our subsequent analysis considers gun ownership rate, crime rate, Marijuana legality, and governor’s party by state level. The frequent pattern uses FP-growth [HPY00], and the threshold of minimum support is 50\% of the total transactions of the WP dataset. 

We also apply DBSCAN \cite{khan2014dbscan} to the longitude and latitude of fatal police shooting locations to identify hotspot clusters. Set parameters eps=0.5 and min\_sample=50, we find the dense areas of fatal police shootings, see below Figure-11. We discover that Los Angles and Atlanta metropolitan areas have two of the largest hotspots. Generally, all the fatal police shooting hotspots are in the top population cities in the country.

\begin{figure}[h!]
  \includegraphics[scale=0.25]{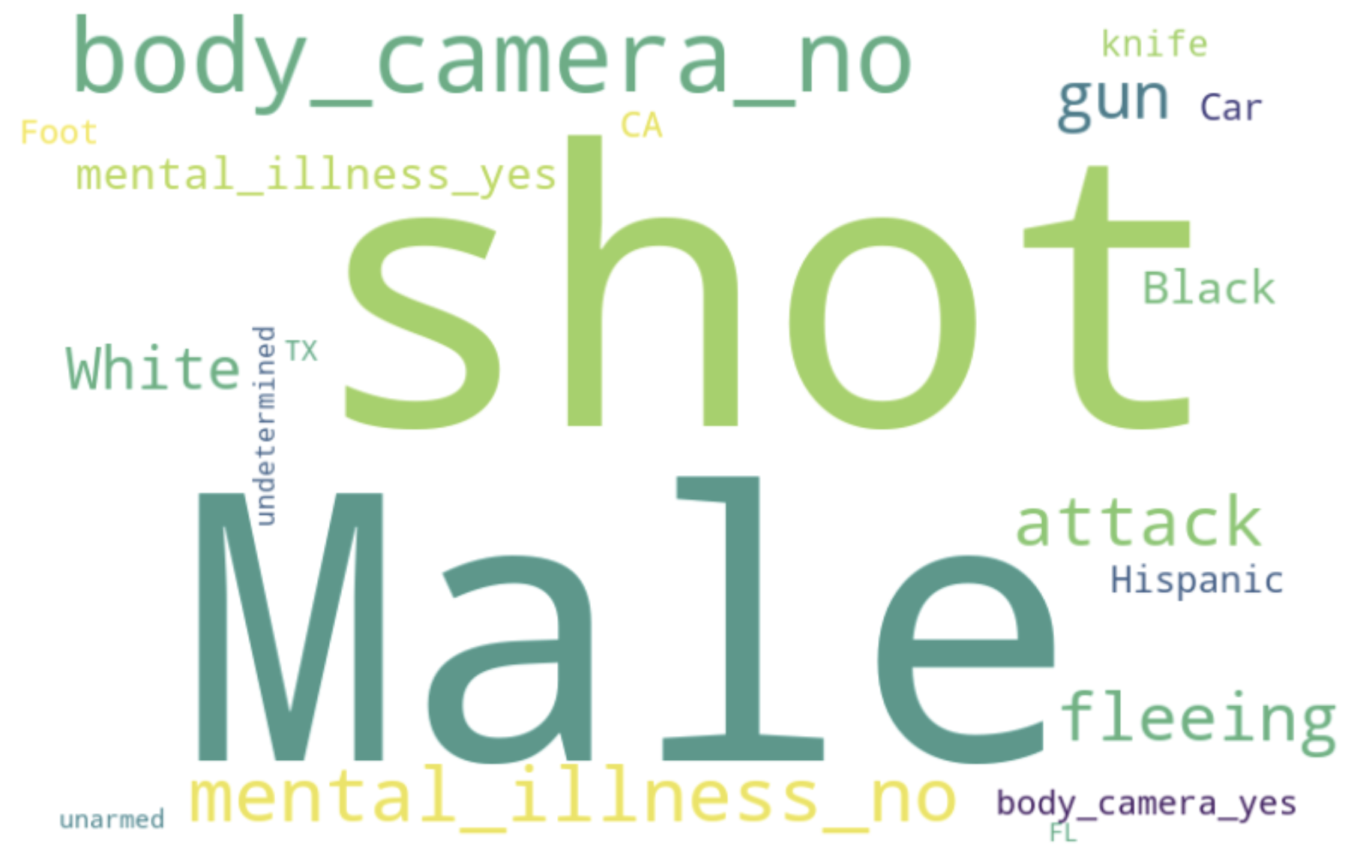}
  \caption{Word cloud of police shooting}
\end{figure}

\begin{figure}[h!]
  \centering
  \includegraphics[scale=0.4]{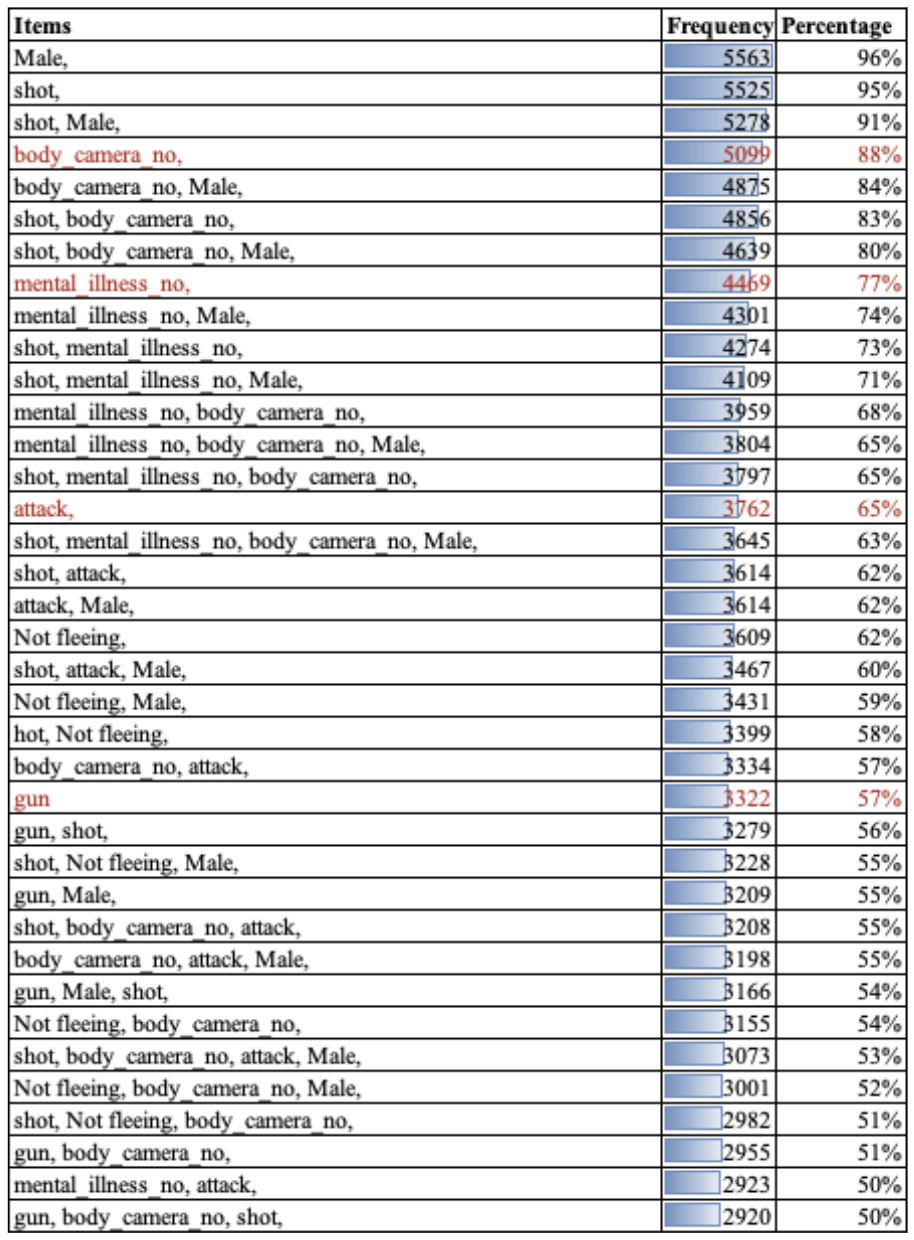}
  \caption{Frequent pattern of police shooting}
\end{figure}

\begin{figure}[h!]
  \includegraphics[scale=0.25]{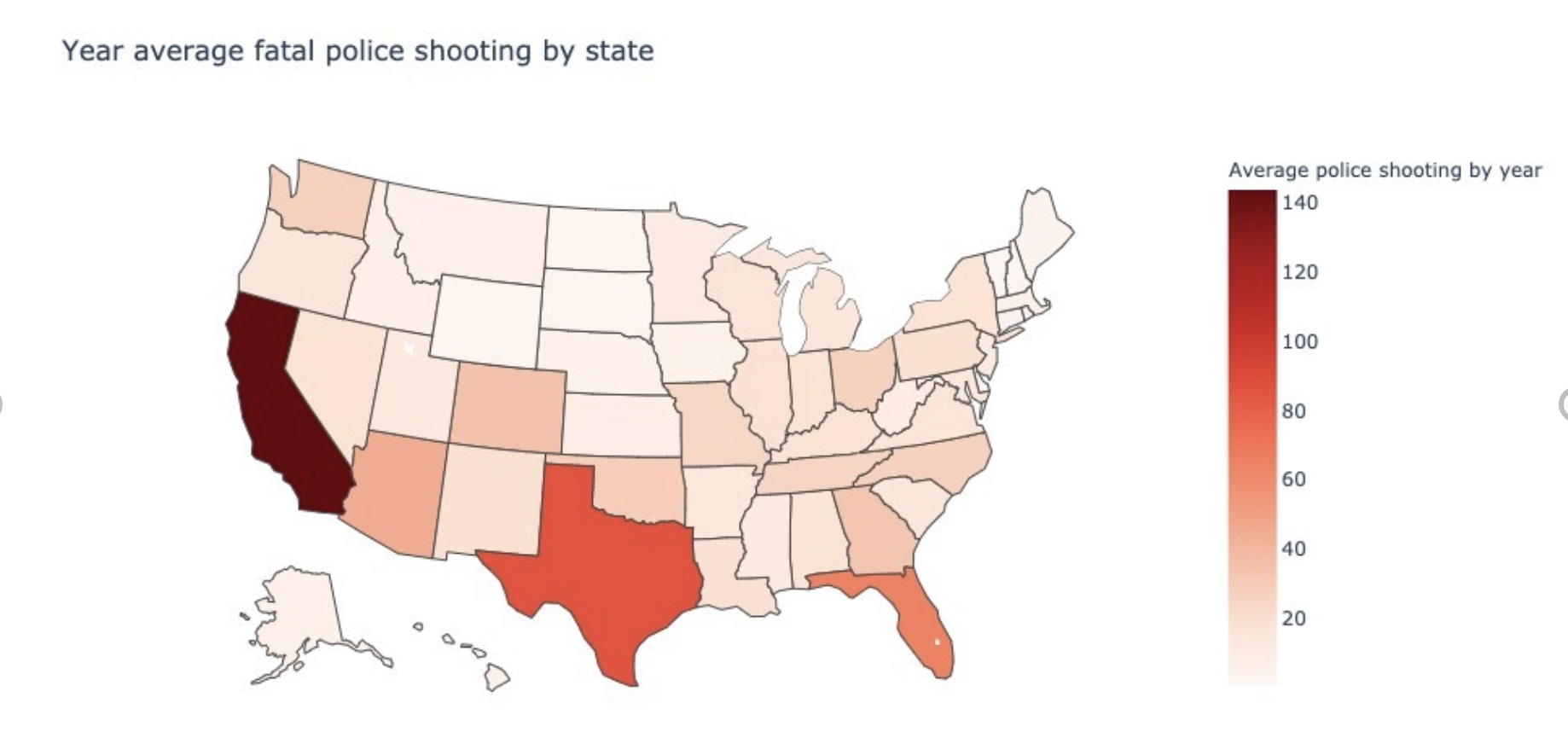}
  \caption{Yearly average Fatal Police shooting per 1m by State}
\end{figure}

\begin{figure}[h!]
  \includegraphics[scale=0.27]{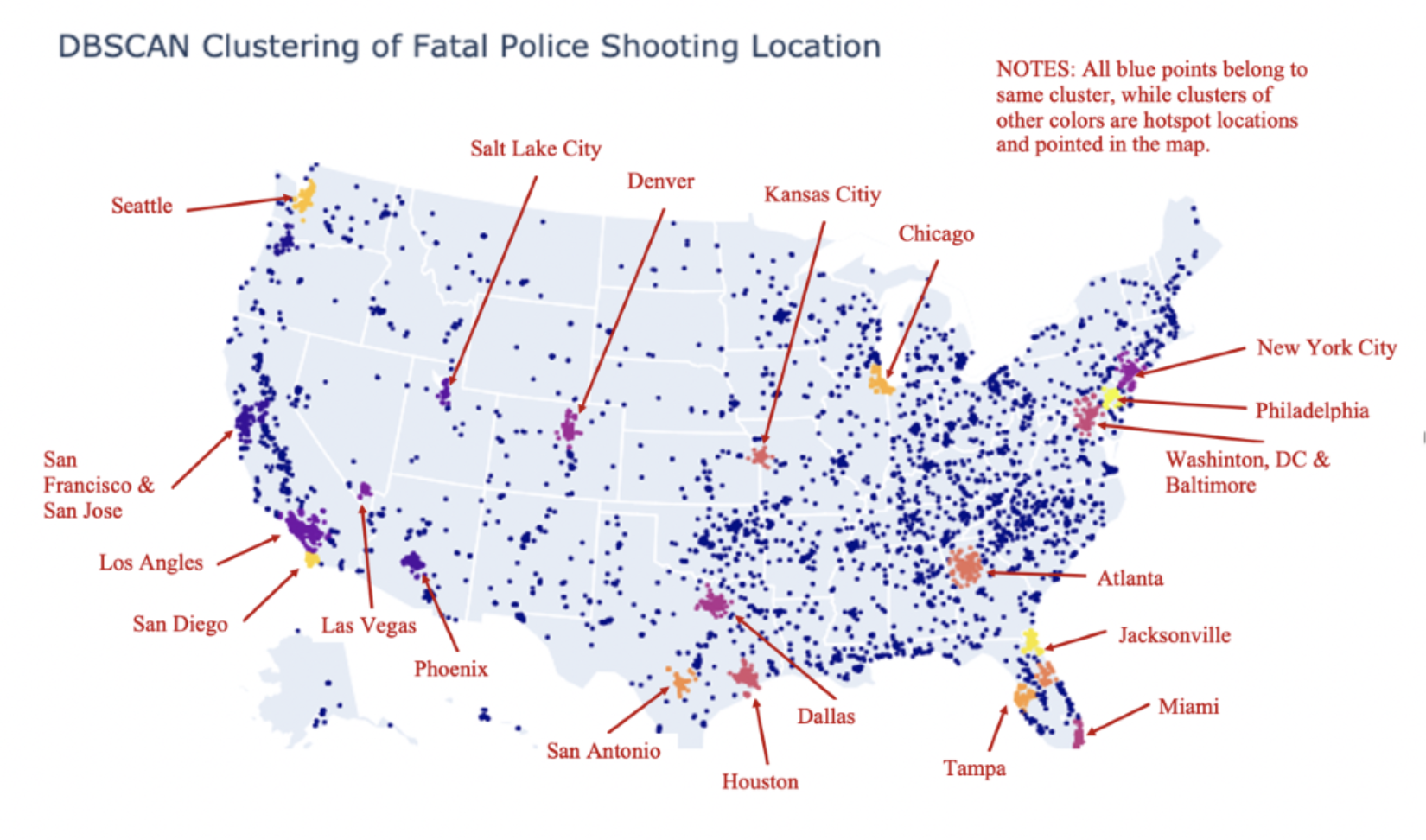}
  \caption{Fatal police shooting hotspots distribution}
\end{figure}

\subsection{Correlated variables analysis}
\subsubsection{Quantitative Variable analysis }
To avoid the population distorting the analysis, we normalized the number to the yearly average fatal police shooting per one million people (\textbf{fatal police shooting rate}). We use this density-kind value for the analysis afterwards. Figure-12 shows that every year on average, how many people were shot by police. \textbf{New Mexico} and \textbf{Alaska} where have relatively less population, become the top state. The color is getting darker from east to west except for large population states such as California, Washington. Doesn't it look like the U.S. history of territory expansion? 
\begin{figure}[h!]
  \includegraphics[scale=0.25]{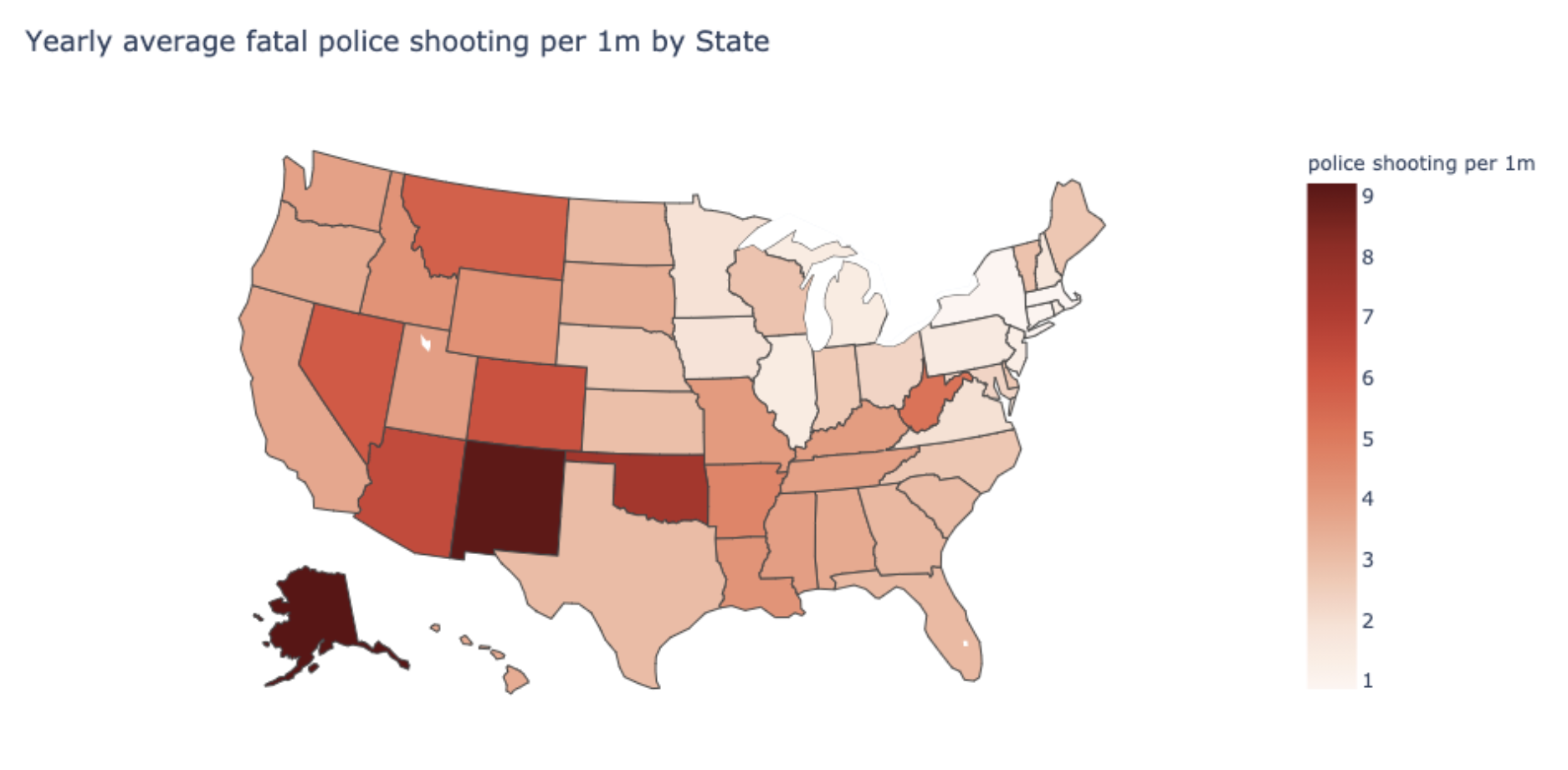}
  \caption{Yearly average Fatal Police shooting per 1m by State}
\end{figure}

It looks the longer the state joined the U.S., the lower the fatal police shooting rate in that state. The correlation coefficient is 68\% between the fatal police shooting rate and the U.S. history of territory expansion. Our interpretation is:  the reason that U.S. police use excessive violence may root from the westward expansion when handling the violent criminals, see Figure-13. 

\begin{figure}[h!]
  \includegraphics[scale=0.25]{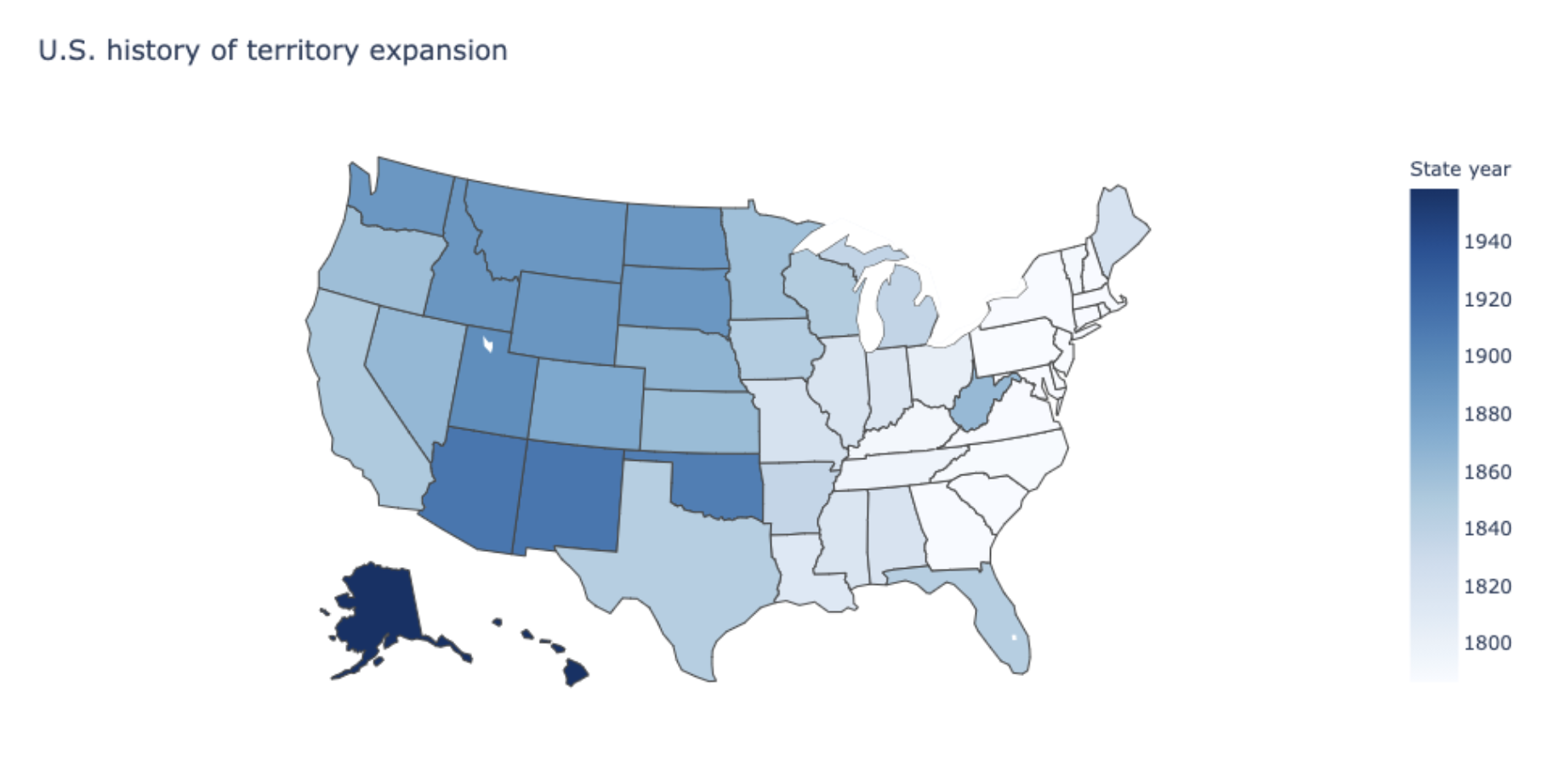}
  \caption{US history of territory expansion}
\end{figure}

The correlation coefficient is 64\% between gun ownership rate \cite{2020Gun} and the fatal police shooting rate. 57\% of victims hold guns (not including other weapons), and 65\% of victims chose to attack police. This hold gun rate doubles than the average gun ownership rate among the country, which is 30\% according to Pew’s report \cite{2020PewResearch}, see Figure-14.  

\begin{figure}[h!]
  \includegraphics[scale=0.25]{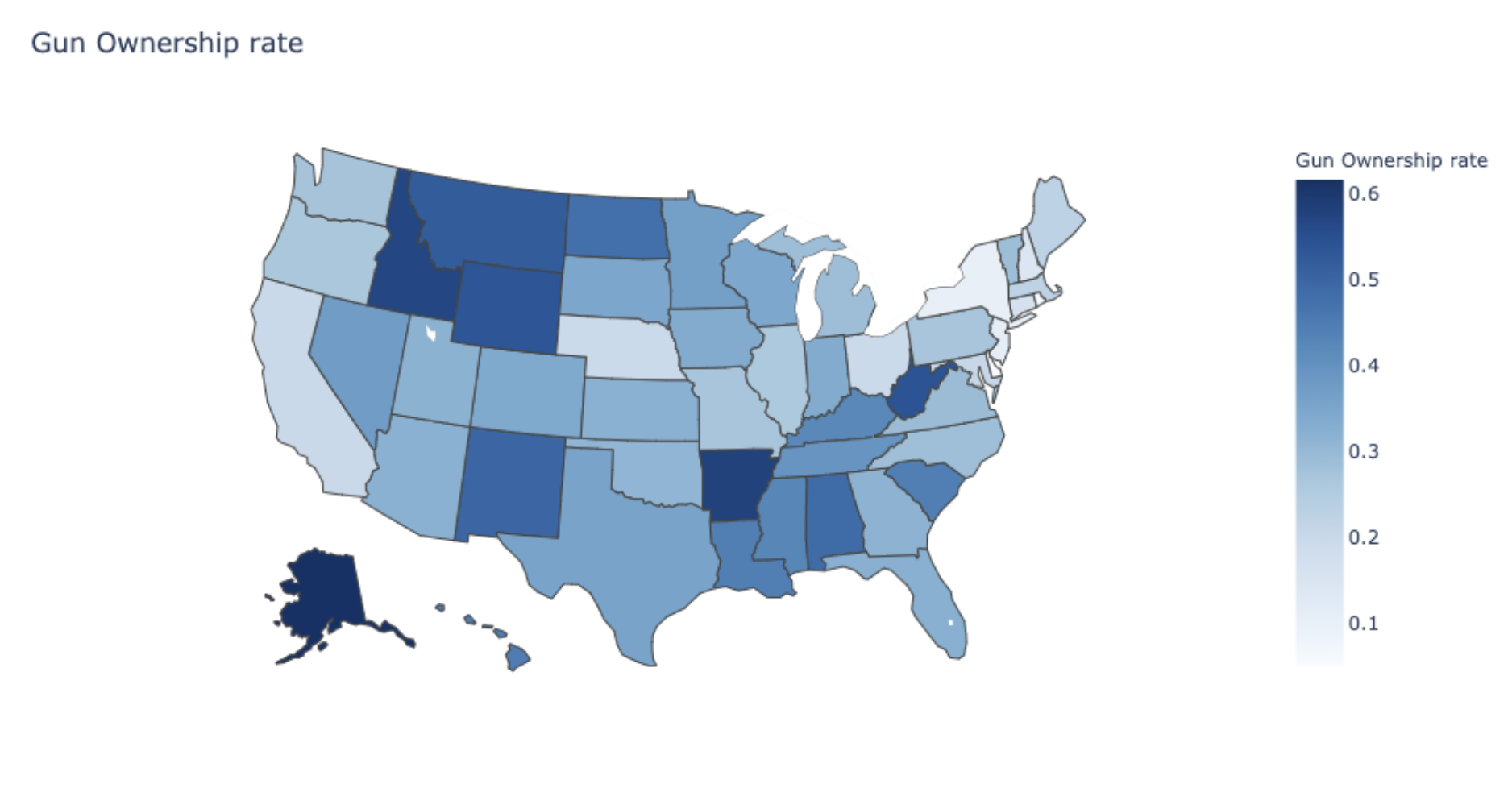}
  \caption{Gun ownership rate by state}
\end{figure}

\begin{figure}[h!]
  \includegraphics[scale=0.25]{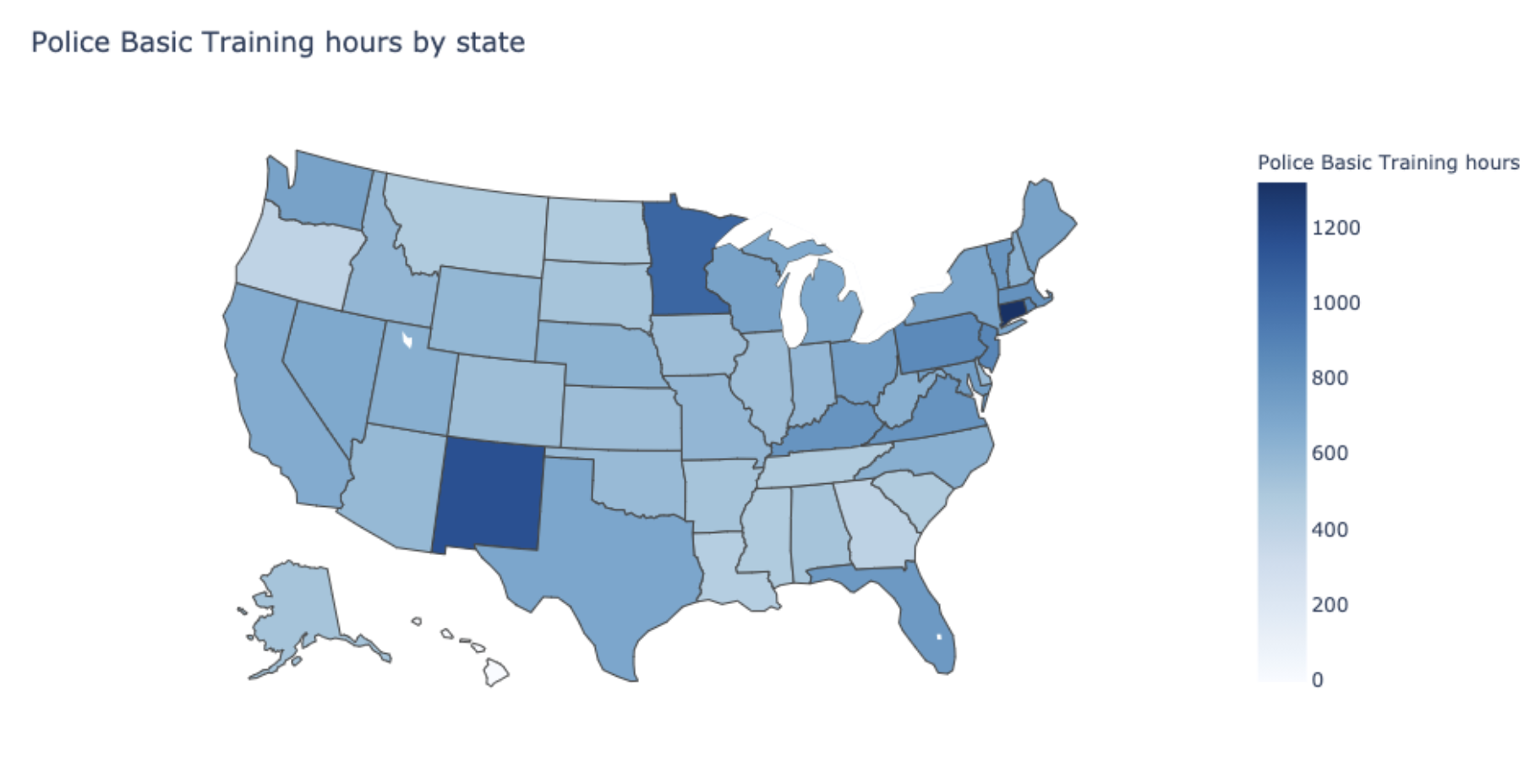}
  \caption{Police Basic Training hours by state}
\end{figure}

The third high correlation variable is the land area \cite{US_states_area}, 59\%, followed by violence rate \cite{StateCrimeRate}, 48\%, poverty rate 37\%, unemployment rate 29\%, see Figure-16. Surprisingly, police basic training hours negatively correlate with the fatal police shooting rate, see Figure-15. Although TrainingReform \cite{2020TrainingReform} appeals appeal to increase police training hours, the current data shows the opposite result. It may suggest reviewing and improving the training itself rather than a single slogan for more hours. 

\begin{figure}[h!]
  \includegraphics[scale=0.25]{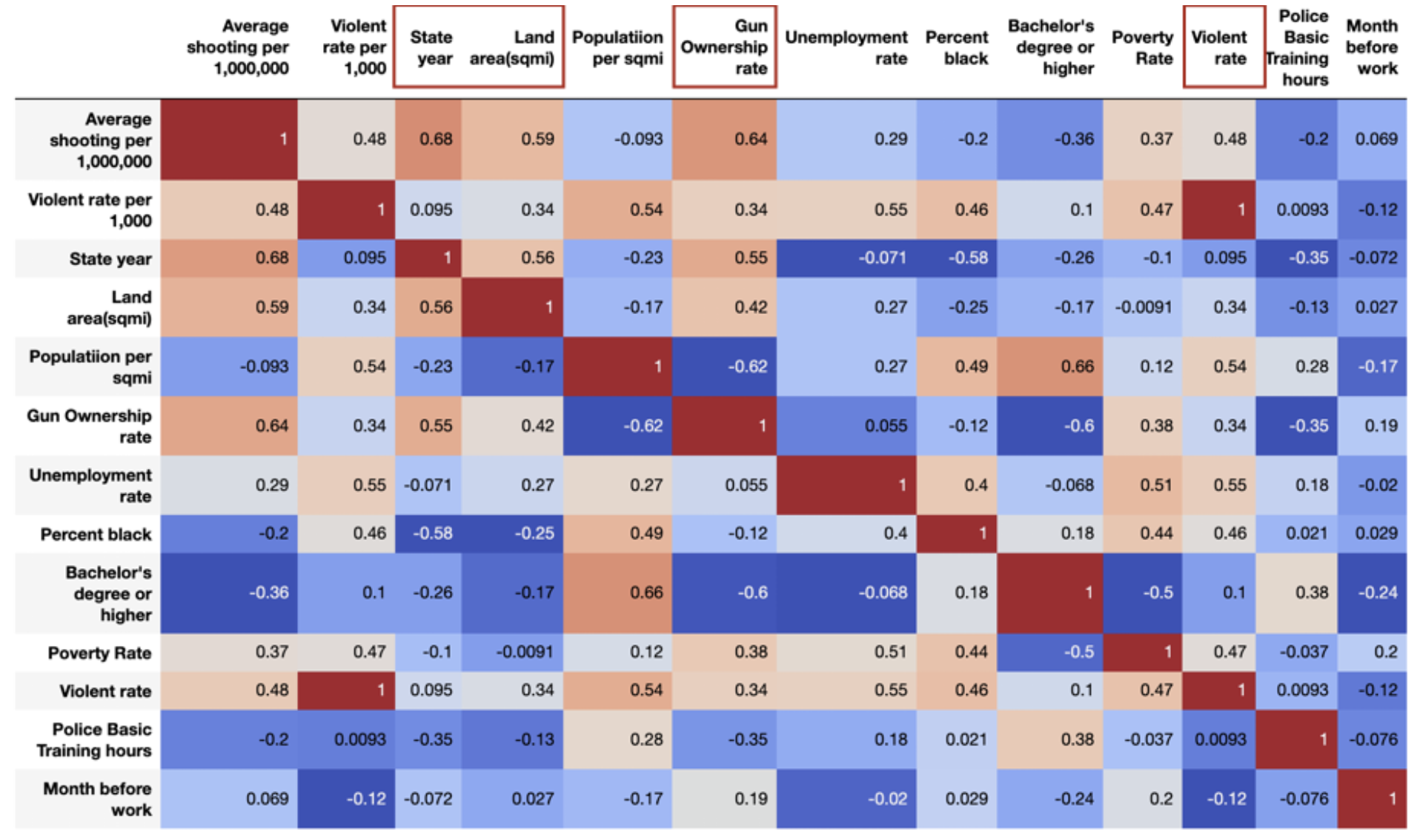}
  \caption{Correlation table}
\end{figure}

We also tested the correlation coefficient's significance to guarantee the association, which are all proved with relatively small p\_value, see Table-1.  
$$t = r\sqrt\frac{n-2}{1-r^2},  \alpha = 0.01$$

\begin{table}[h!]
  \includegraphics[scale=0.3]{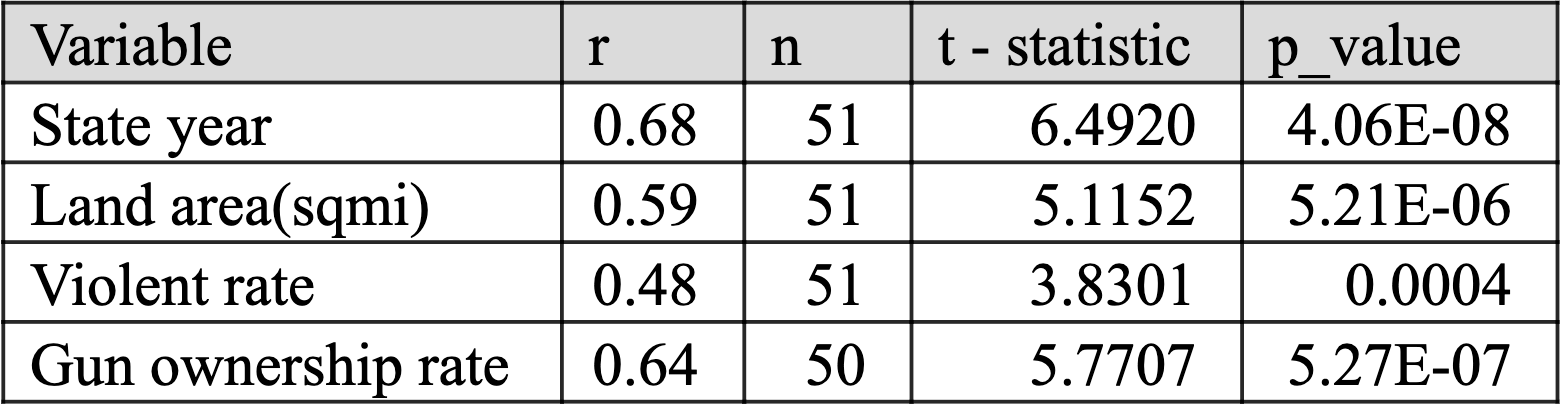}
  \caption{Correlation coefficient test }
\end{table}

\subsubsection{Categorical variables analysis}
In this part, we tested the significance of the fatal police shooting rate by state-level Governor’s party \cite{US_Governors} and Marijuana Legality \cite{2020Marijuana}. We failed to reject the null hypothesis, and we can conclude that there is no difference between those states on the fatal police shooting. 

\begin{figure}[h!]
  \includegraphics[scale=0.35]{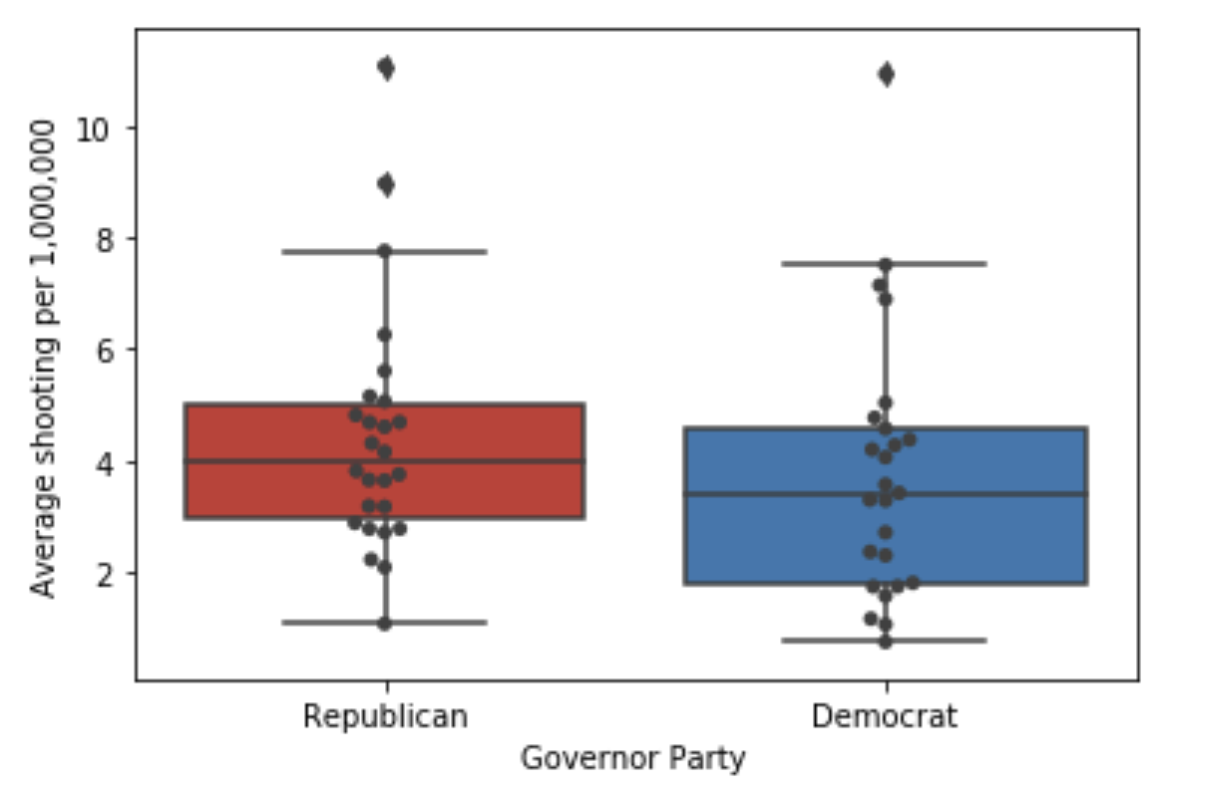}
  \caption{Boxplot of fatal police shooting in Republican and Democrat states}
\end{figure}

T-test results:

$H_0$: $\mu_{GOP} = \mu_{Dems}$

$H_A$: the average fatal police shooting rate are not equal between Republican and Democrat governor states 

Test result: 
since ${value}$ = 0.3254 > 0.05, we failed to reject the null hypothesis. The average fatal police shooting rate are equal between Republican and Democrat governor states

\begin{figure}[h!]
  \includegraphics[scale=0.35]{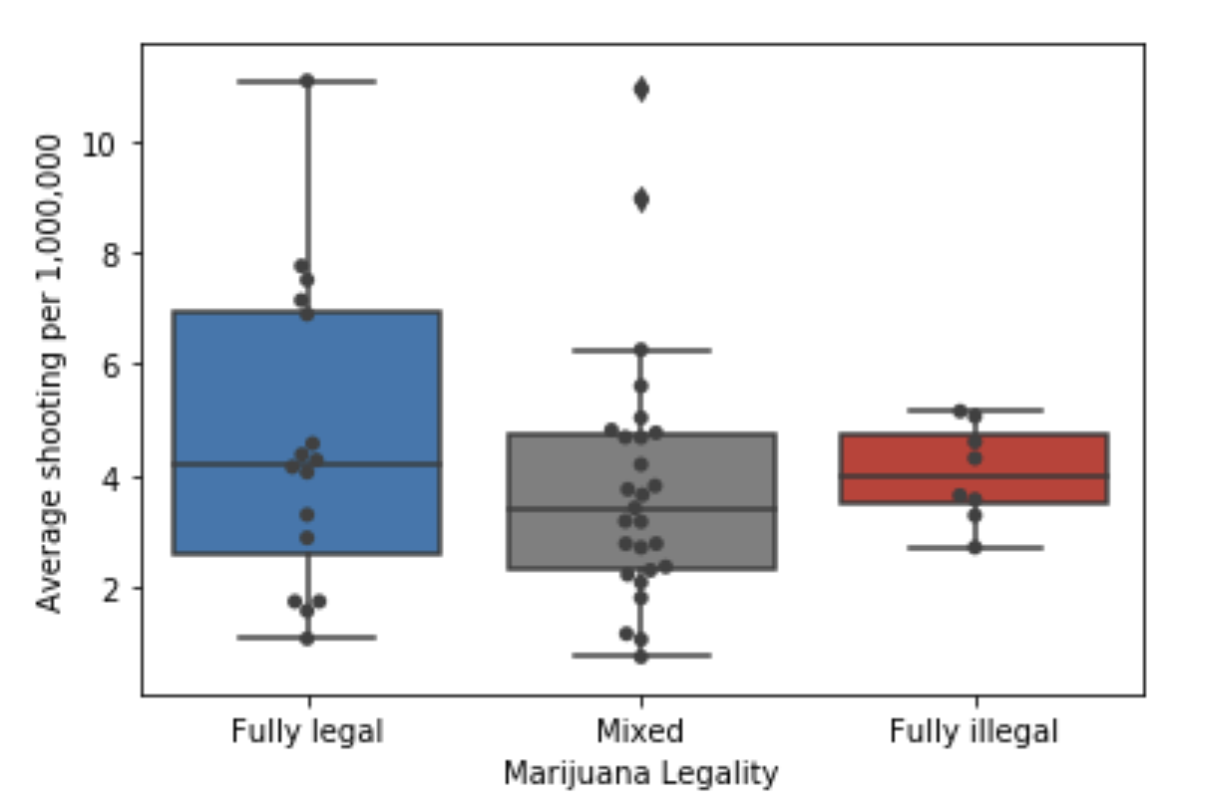}
  \caption{Boxplot of fatal police shooting among different marijuana legality states}
\end{figure}

One-way ANOVA: 

$H_0$: $\mu_{FL} = \mu_{MML} = \mu_{FI}$

$H_A$: at least one of the average rates differs from one of the others 

Test result:  
F = 0.6492, $P_{value}$ = 0.527, fail to reject the null hypothesis. The average fatal police shooting rate are equal among different marijuana legality states

\section{Fatal police shooting rate and victims race prediction}
In this part, we used the insights we draw from WP data and multi-attributes correlation analysis to build predictive models. We constructed a series of regression models to predict fatal police shooting rates on the state level and a series of classification models to predict fatal police shooting victims' race.  

\subsection{Fatal police shooting rate prediction on state level}
According to above correlation analysis, we chose \textbf{the violent crime rate}, \textbf{land area}, and \textbf{gun ownership rate}, \textbf{state\_joined\_year} based on their highest correlation coefficient with the fatal police shooting rate. We acquired more data points by looking at each state every year from 2015 to 2019 separately.

In the Weka machine learning software, we tried all models and chose three of the best-performed models based on ten-fold cross-validation performance. The best one is Kstar \cite{cleary1995k}. It achieved 28.04\% cross-validation relative absolute error and explained 88.53\% variance, followed by K-Nearest-Neighbor Regression and Random Forest. These three models all performed much better than the baseline linear regression model, see Table-2. 

\begin{table}[h!]
  \includegraphics[scale=0.22]{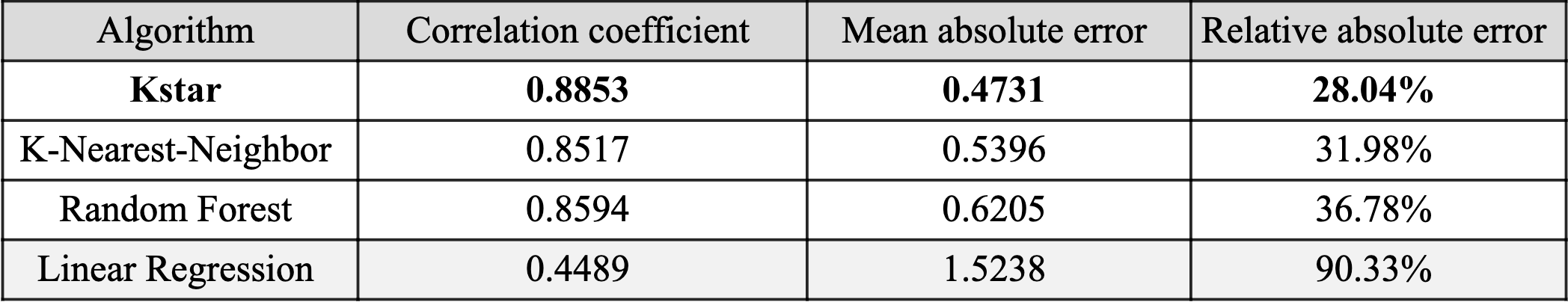}
  \caption{Ten-fold cross validation results}
\end{table}

Figure-19 displays the cross-validation prediction error of each data point in the Kstar model (each data point represents the fatal police shooting rate of a state in a particular year). The X-axis is the real police shooting rate, while the Y-axis is the predicted police shooting rate. The large cross means a higher error rate. 

\begin{figure}[h!]
  \includegraphics[scale=0.24]{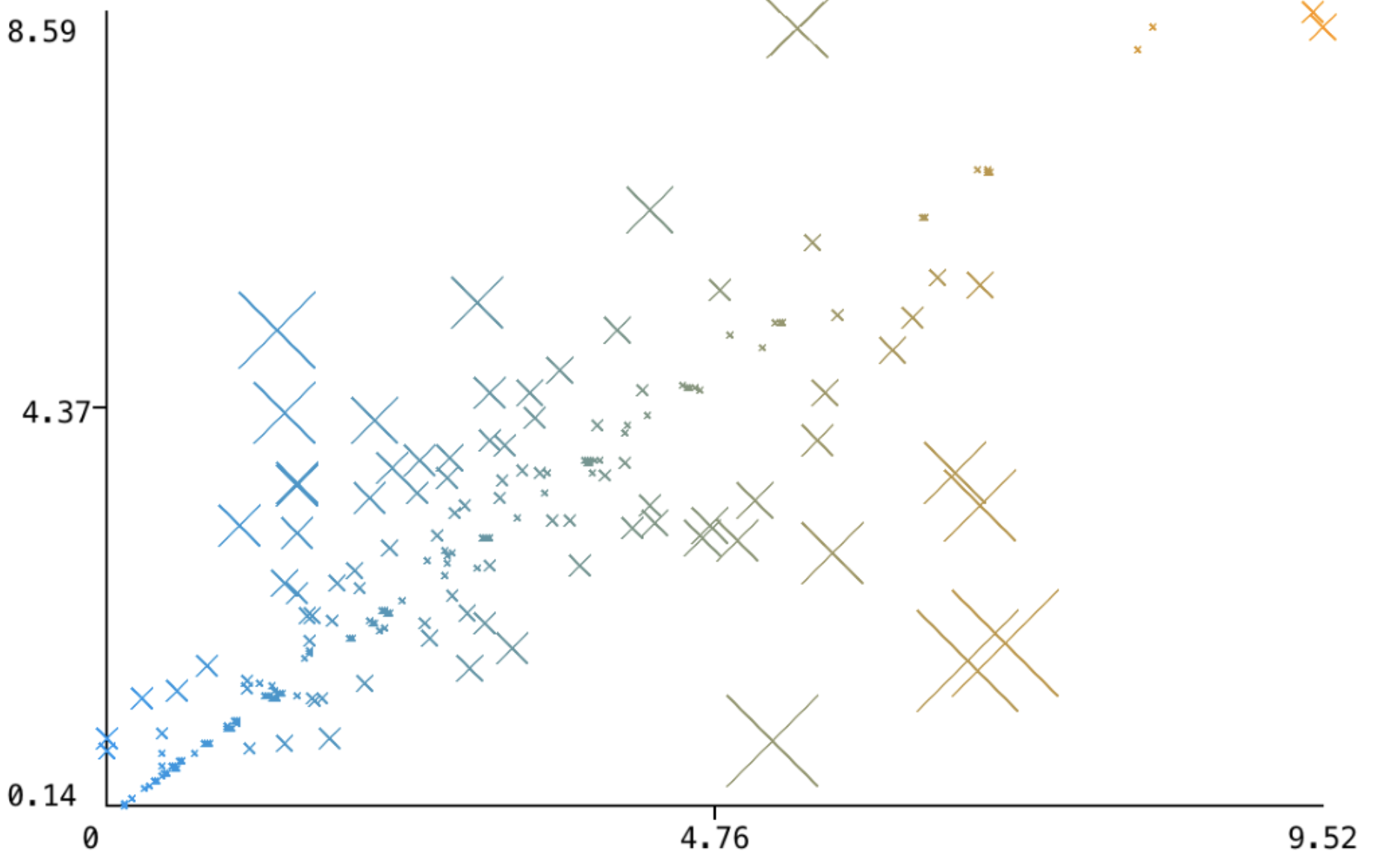}
  \caption{Predicted fatal police shooting rate vs. Real fatal police shooting rate}
\end{figure}

The prediction model tells us that the reason for fatal police shootings could be complex. It is related to the state joined year, state land area, gun ownership rate, and violent crime rate. It suggests us to understand this problem from multi-dimensional aspects. 

\subsection{Predict victims’ race in fatal police shooting}
This prediction intends to test whether or not there is racial discrimination during the fatal police shooting. The null hypothesis is that the model cannot predict the victim’s race (No racial discrimination). The alternative hypothesis is that the model can predict the victim’s race (racial discrimination). We use WP data from 01/01/2015 to 02/12/2020 and excluded the data missing the race information. The total records are 4518. Since “age” is the only numeric variable, we applied the chi-square test to select the predictor for the rest of the variables. 

\subsubsection{Chi-square testing}
$$\chi^2 = \sum\frac{(O_i - E_i)^2}{E_i}\\,  \alpha = 0.05$$
where $\chi^2$ = chi squared, $O_i$ = observed value, $E_i$ = expected value

\begin{table}[h!]
  \includegraphics[scale=0.2]{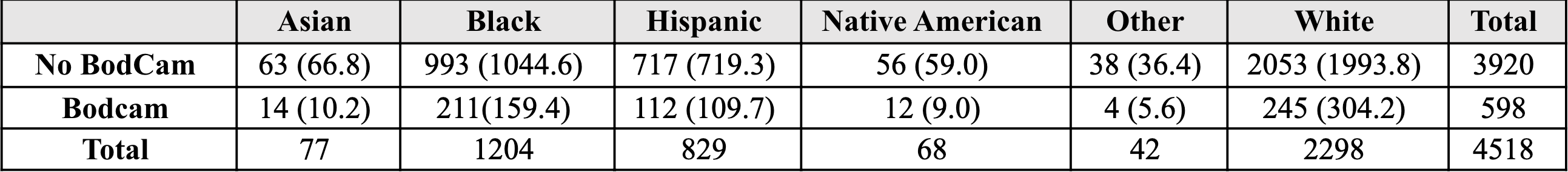}
  \caption{The chi\_square contingency table for body\_ camera}
\end{table}

\begin{table}[h!]
  \includegraphics[scale=0.2]{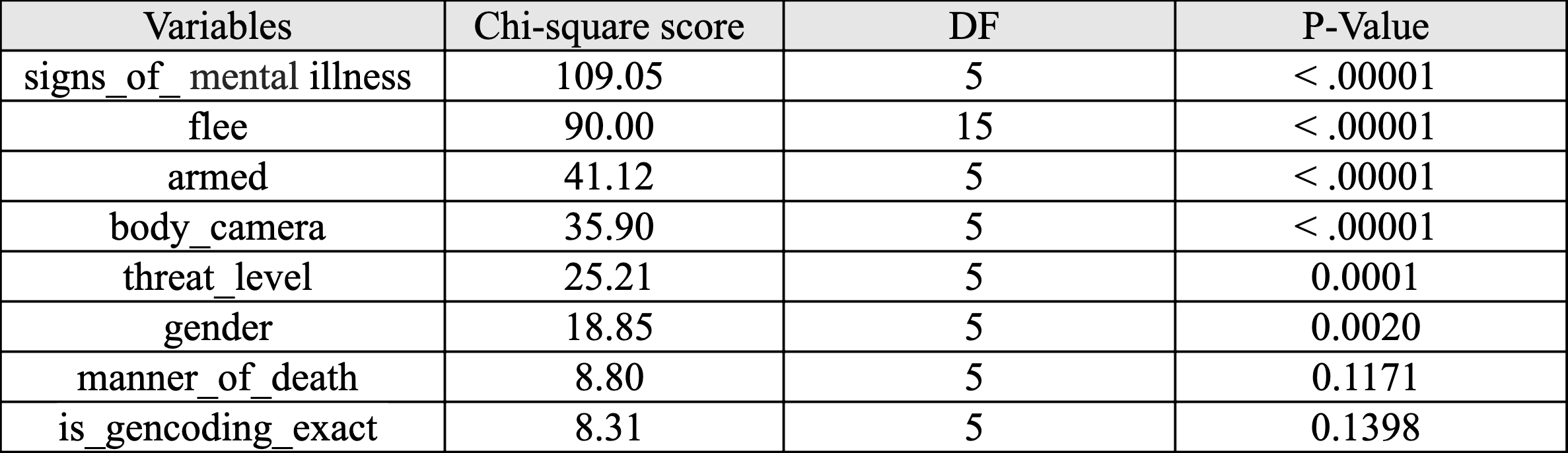}
  \caption{Chi-square testing for categorical variables}
\end{table}

After applying chi-square testing to the above categorical variables, we find that threat\_level, signs\_of\_mental\_illness, armed, flee, body\_camera, and gender are not independent of the race at 0.05 statistically significant level, see Table 4. On the other hand, manner\_of\_death and is\_gencoding\_exact are independent of the race at 0.05 statistically significant level. For city and state, the degree of freedoms (DF) is too large to apply chi-square testing. Finally, we chose \textbf{armed}, \textbf{age}, \textbf{gender}, \textbf{signs\_of\_mental\_illness}, \textbf{threat\_level}, \textbf{flee}, and \textbf{body\_camera} as predictors and city, age as back-up predictors for the racial classification model. 

\subsubsection{Classification model}
In the Weka machine learning software and Python AutoML package, we tried all models and chosen the top three best-performed models based on stratified five-fold cross-validation performance. see Table-5 below. 

\begin{table}[h!]
  \includegraphics[scale=0.28]{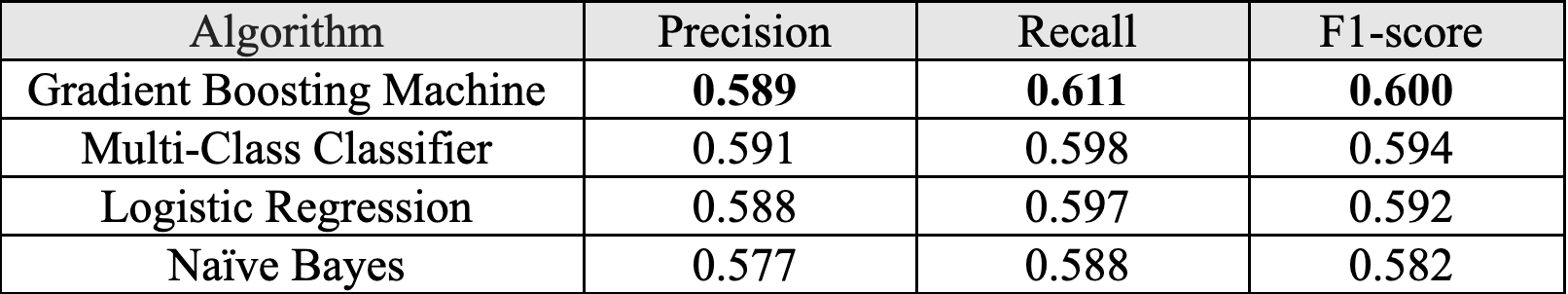}
  \caption{Stratified cross validation results}
\end{table}
 
We find that adding city and state attributes could boost model performance. Gradient Boosting Machine \cite{friedman2001greedy} performs best, having 0.589 precision and 0.611 recall, slightly better than predicting all victims to be white (about 50\% precision and recall). GBM algorithm gives us an idea of the importance of attributes we selected for prediction. \textbf{City}, \textbf{state}, \textbf{armed}, and \textbf {age} attributes play essential roles in racial prediction. See Figure-20 below. We failed to reject the null hypothesis since even the best-performed model cannot predict victims’ race well, proving that there is no racial discrimination for observed fatal police shootings in WP data. 

\section{Conclusion}
In conclusion, we found that mainstream media disproportional reporting fatal police shooting by the race, which may instigate hostile sentiments between police and the public. We suggest mainstream media report all news according to the realistic. Second, we found that the police shooting rate depends on many variables. The top four significant attributes were \textbf{state joined year}, \textbf{state land area}, \textbf{gun ownership rate}, and {violent crime rate}. Choosing these four attributes as predictors, our best-performed regression model could predict the fatal police shooting rate with about 88.53\% correlation coefficient. Admittedly, we cannot find all the influence factors. It indicates that the fatal police shooting is a \textbf{complex multi-dimensional} problem. We also found two variables (police basic training hour, number of months police can work before basic training) appealed by CNBC negatively and weakly correlated with the fatal police shooting. Third, based on the WP dataset, we tried to depict a typical scenario when a police shooting happened and remark the hotspots among the country. Last, our three best performance models show no significant evidence to conclude that racial discrimination happened during fatal police shootings recorded by the WP dataset. 

{\small
\bibliographystyle{ieee_fullname}
\bibliography{egbib}

\begin{thebibliography}{10}\itemsep=-1pt

\bibitem{cleary1995k}
John~G Cleary and Leonard~E Trigg.
\newblock K*: An instance-based learner using an entropic distance measure.
\newblock In {\em Machine Learning Proceedings 1995}, pages 108--114. Elsevier,
  1995.

\bibitem{2020Marijuana}
{DISA Global Solutions}.
\newblock {Map of Marijuana Legality by State}.
\newblock \url{https://disa.com/map-of-marijuana-legality-by-state}.

\bibitem{StateCrimeRate}
{FBI}.
\newblock {Crime in the United States, by Region, Geographic Division, and
  State}.
\newblock
  \url{https://ucr.fbi.gov/crime-in-the-u.s/2017/crime-in-the-u.s.-2017/topic-pages/tables/table-4}.

\bibitem{friedman2001greedy}
Jerome~H Friedman.
\newblock Greedy function approximation: a gradient boosting machine.
\newblock {\em Annals of statistics}, pages 1189--1232, 2001.

\bibitem{khan2014dbscan}
Kamran Khan, Saif~Ur Rehman, Kamran Aziz, Simon Fong, and Sababady Sarasvady.
\newblock Dbscan: Past, present and future.
\newblock In {\em The fifth international conference on the applications of
  digital information and web technologies (ICADIWT 2014)}, pages 232--238.
  IEEE, 2014.

\bibitem{2020KBP}
{KilledByPolice}.
\newblock {Police Shootings Database – Killed By Police}.
\newblock \url{https://killedbypolice.net/}.

\bibitem{2020PewResearch}
{Kim Parker, Juliana Menasce Horowitz, Ruth Igielnik, J. Baxter Olophant and
  Anna Brown}.
\newblock {The demographics of gun ownership}.
\newblock
  \url{https://www.pewsocialtrends.org/2017/06/22/the-demographics-of-gun-ownership/}.

\bibitem{mentch2020racial}
Lucas Mentch.
\newblock On racial disparities in recent fatal police shootings.
\newblock {\em Statistics and Public Policy}, 7(1):9--18, 2020.

\bibitem{ross2015multi}
Cody~T Ross.
\newblock A multi-level bayesian analysis of racial bias in police shootings at
  the county-level in the united states, 2011--2014.
\newblock {\em PloS one}, 10(11):e0141854, 2015.

\bibitem{2020TrainingReform}
{The Institute for Criminal Justice Training Reform}.
\newblock {State Law Enforcement Training Requirements}.
\newblock \url{https://www.trainingreform.org/}.

\bibitem{1776Thomas}
{Thomas Jefferson}.
\newblock {United States Declaration of Independence}.
\newblock
  \url{https://en.wikipedia.org/wiki/United_States_Declaration_of_Independence}.

\bibitem{CrimeRatebyRace}
{U.S. Department of Justice}.
\newblock {Criminal Victimization}.
\newblock \url{https://www.bjs.gov/content/pub/pdf/cv19.pdf}.

\bibitem{2020WashingtonPost}
{Washington Post}.
\newblock {Fatal Police Shooting Dataset}.
\newblock \url{https://github.com/washingtonpost/data-police-shootings}.

\bibitem{2020WP}
{Washington Post}.
\newblock {Police Shootin Database}.
\newblock
  \url{https://www.washingtonpost.com/graphics/investigations/police-shootings-database/}.

\bibitem{US_Governors}
{Wikipedia}.
\newblock {List of current United States governors}.
\newblock
  \url{https://en.wikipedia.org/wiki/List_of_current_United_States_governors}.

\bibitem{US_states_area}
{Wikipedia}.
\newblock {List of U.S. states and territories by area}.
\newblock
  \url{https://en.wikipedia.org/wiki/List_of_U.S._states_and_territories_by_area}.

\bibitem{2020Gun}
{World Population Review}.
\newblock {Gun Ownership by State 2020}.
\newblock
  \url{https://worldpopulationreview.com/state-rankings/gun-ownership-by-state}.

\end{thebibliography}
}

\end{document}